\documentclass[prl,aps,twocolumn,showpacs,groupedaddress]{revtex4-1}

\usepackage[centertags]{amsmath}
\usepackage{graphicx}
\usepackage{amsfonts}
\usepackage{epstopdf}
\usepackage{ulem}
\usepackage{color}

\begin{document}

\title{Time-Reversal Generation of Rogue Waves}

\author{Amin Chabchoub$^{1}$ and Mathias Fink$^2$}

\affiliation{$^1$ Centre for Ocean Engineering Science and Technology, Swinburne University of Technology, Hawthorn, Victoria 3122, Australia}
\email{achabchoub@swin.edu.au}
\affiliation{$^2$ Institut Langevin, ESPCI \& CNRS, UMR CNRS 7587, 1 rue Jussieu, 75005 Paris, France}

\begin{abstract}
The formation of extreme localizations in nonlinear dispersive media can be explained and described within the framework of nonlinear evolution equations, such as the nonlinear Schr\"odinger equation (NLS). Within the class of exact NLS breather solutions on finite background, which describe the modulational instability of monochromatic wave trains, the hierarchy of both in time and space localized rational solutions are considered to be appropriate prototypes to model rogue wave dynamics. Here, we use the time-reversal invariance of the NLS to propose and experimentally demonstrate a new approach to construct strongly nonlinear localized waves focused both in time and space. The potential areas of applications of this time-reversal approach range from remote sensing to motivated analogous experimental analysis in other nonlinear dispersive media, such as optics, Bose-Einstein condensates and plasma, where the wave motion dynamics is governed by the NLS. 
\end{abstract}
\maketitle

Rogue waves have received considerable interest recently \cite{KPS,Osborne,OnoratoReport,Toffoli,KiblerKM,Guoa,ChabchoubPRX}. The sudden formation of extreme waves in the ocean is well reported, and no longer doubted in the scientific community \cite{KPS}. Besides the trivial linear superposition principle of waves, one possible mechanism explaining the formation of rogue waves, characterized being strongly localized, is the modulational instability of weakly nonlinear monochromatic waves, first discovered in water waves \cite{BF}. This instability can be modeled within the framework of the nonlinear Schr\"odinger equation (NLS) \cite{BennyNewell,Zakharov}, an evolution equation, which describes the dynamics in time and space of wave trains in waters of finite and infinite depth \cite{Osborne}. Within the class of exact breather solutions on finite background \cite{Kuznetsov,Akhmediev} there is a hierarchy of both in time and space localized solutions \cite{Peregrine,Akhmediev2,Akhmediev3}, which amplify the amplitude of the carrier by a factor of three and higher. Owning these properties, the latter are considered to be appropriate solutions to describe the formation of rogue waves \cite{Shrira,OnoratoReport}. 

Recent observations of these doubly-localized NLS solutions in optics \cite{Kibler}, in water waves \cite{ChabchoubPRL,ChabchoubPRE,OnoratoPlosOne} and in plasma \cite{Bailung} confirmed the ability of the NLS to model strong localizations in nonlinear dispersive media and justified the choice of the NLS approach. In this paper, we study the implication of the time-reversal invariance of the NLS equation and we propose a new way to experimentally focus both in time and space rogue waves using the principle of the time-reversal mirrors that was first extensively studied for acoustic and elastic waves \cite{Fink1}. In a standard time-reversal experiment, the wave field radiated by a source is first measured by an array of antennas positioned in the far field of the source and then time-reversed and simultaneously rebroadcasted by the same antenna array. Due to the time-reversal invariance of the wave process, the reemitted wave field focuses back in space and time on the original source, whatever the complexity of the propagation medium. The effect of dispersion \cite{Fink2,Fink3} and  nonlinearities \cite{Fink4} has been experimentally studied for acoustic waves, and it has been shown that the time-reversed field focuses back in time and space as long as nonlinearities do not create dissipation, i.e., as long as the propagation distance is smaller than the shock distance. Time-reversal in water waves has also been recently confirmed \cite{Fink5}, however, the nonlinear effects were negligeable in this last study. 

In this letter, we confirm experimentally the refocusing of the time-reversed field of a doubly-localized NLS breather rogue wave solutions, related to the modulational instability. Our results are in excellent agreement with theory and show that the latter technique can be applied to nonlinear waves, which propagate in a wide range of nonlinear dispersive systems described by the NLS, and may be used to construct new strongly localized breather-type solutions as well as to analyze and predict rogue wave dynamics.

The evolution dynamics in time and space of nonlinear wave trains in deep-water can be modeled using the focusing NLS \cite{Zakharov,Ono}, given by: 
\begin{equation}\label{nls}
i\left(\frac{\displaystyle\partial A}{\displaystyle \partial
t}+\frac{\displaystyle\omega_0}{\displaystyle2k_0}\frac{\displaystyle\partial A}{\displaystyle \partial
x}\right)-\frac{\displaystyle\omega_0}{\displaystyle
8k_0^2}\frac{\displaystyle\partial^2 A}{\displaystyle \partial
x^2}-\frac{\displaystyle\omega_0 k_0^2}{\displaystyle
2}\left|A\right|^2A=0,
\end{equation}
while the free water surface elevation to first order in steepness is: 
\begin{equation}
\eta(x,t)=\operatorname{Re}\left(A\left(x,t\right)\cdot\exp\left[i\left(k_0x-\omega_0
t\right)\right]\right).
\label{se}
\end{equation}
Here, $k_0$ and $\omega_0$ are the wave number and wave frequency, respectively. These physical values are connected through the linear dispersion relation of deep-water waves \cite{ChabchoubPRL}. The deep-water NLS coefficients are valid for the condition $k_0h_0\gg 1$, while $h_0$ denotes the water depth. A scaled form of Eq. (\ref{nls}) can easily be obtained by applying straight-forward transformations of the time, space and amplitude variables \cite{ChabchoubPRL}:
\begin{equation}
i\psi_T+\psi_{XX}+2\left|\psi\right|^2\psi=0.
\label{snls}
\end{equation} 
The dimensionless form of the NLS (\ref{snls}) admits an infinite hierarchy of exact breather solutions $\psi_n\left(X,T\right)$ localized both in time and in space. Breathers are pulsating localized wave envelopes, which describe the dynamics of unstable wave trains in nonlinear dispersive media. Doubly-localized breather solutions can be expressed in terms of polynomials $G_n(X,T)$, $H_n(X,T)$ and $D_n(X,T)$:
\begin{align}
\psi_n\left(X,T\right)=&\left(\left(-1\right)^n+\frac{\displaystyle G_n\left(X,T\right)+iH_n\left(X,T\right)}{D_n\left(X,T\right)}\right)\nonumber\\ 
&\times\exp\left(2iT\right),
\end{align}
where $n\in\mathbb{N}$ labels the order of the solution. The lowest and first-order $(n=1)$ solution, is known as the Peregrine breather \cite{Peregrine} and is defined as $G_1=4$, $H_1=16T$ and $D_1=1+4X^2+16T^2.$ , whereas the second-order solution $(n=2)$, referred to as the Akhmediev-Peregrine breather \cite{Akhmediev,Akhmediev2}, is determined by 

\begin{align}
G_2=&-\left(X^2+4T^2+\frac{\displaystyle 3}{\displaystyle 4}\right)\left(X^2+20T^2+\frac{\displaystyle 3}{\displaystyle 4}\right)+\frac{\displaystyle 3}{\displaystyle 4}\\ \label{hos2}
H_2=&2 T\left(3X^2-4T^2-2\left(X^2+4T^2\right)^2+\frac{\displaystyle 15}{\displaystyle 8}\right)\\
D_2=&\frac{\displaystyle 1}{\displaystyle 3}\left(X^2+4T^2\right)^3+\frac{\displaystyle 1}{\displaystyle 4}\left(X^2-12T^2\right)^2
\nonumber\\
&+\frac{\displaystyle 3}{\displaystyle64}\left(12X^2+176T^2+1\right)
\end{align}

\begin{figure}[h]
\centering
\includegraphics[width=\columnwidth]{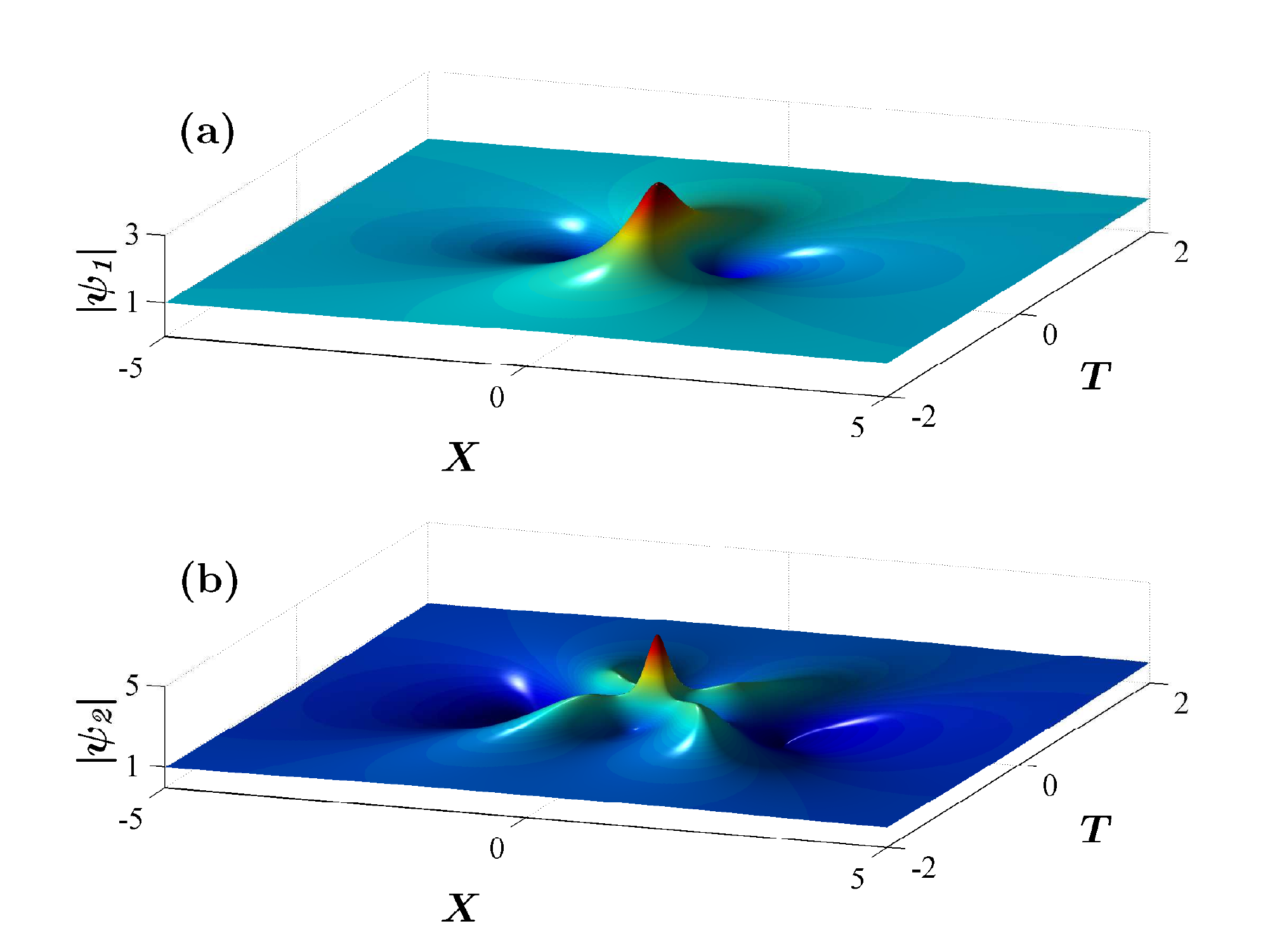}
\caption{(Color online) (a) First-order doubly-localized rational solution (Peregrine breather), which at $X=T=0$ amplifies the amplitude of the carrier by a factor of three. (b) Second-order doubly-localized rational solution (Akhmediev-Peregrine breather), which at $X=T=0$ amplifies the amplitude of the carrier by a factor of five.}\label{fig1}
\end{figure}

The theoretical evolution of these solutions are shown in Fig. \ref{fig1}.

These breathers describe the modulation instability of Stokes waves in the limit of infinite wave modulation period. In particular, they significantly amplify the amplitude of the carrier and therefore, increase the nonlinearity during the evolution of the monochromatic wave field. Generally, the amplitude amplification factor of the $n$-th order rational solution is at $X=T=0$ of $2n+1$. Hence, localized waves modeled by higher-order breathers are also called super rogue waves. The Peregrine breather has been recently observed in several nonlinear dispersive media \cite{Kibler,ChabchoubPRL,Bailung}, whereas up to now, the second-order solution has only been observed in water waves \cite{ChabchoubPRX}. These observations confirm the ability of the NLS to model extreme wave localizations, which naturally engender a severe broadening of the spectrum during their evolution.

As the time-dependent part of the NLS equation, contains a term in $i\frac{\displaystyle\partial A}{\displaystyle \partial
t}$, it is easy to see that if $A\left(x,t\right)$ is a solution of Eq. (\ref{nls}), then $A^*\left(x,-t\right)$ is also a solution. Therefore, both $\eta\left(x,t\right)$ and $\eta\left(x,-t\right)$ describe two possible solutions for the water surface elevation. Thanks to this property, we can use a time-reversal mirror to create the time-reversed wave field $\eta\left(x,-t\right)$ in the whole propagating medium. It is sufficient for the one-dimensional problem to measure the wave field $\eta\left(x,t\right)$ at one unique point $x_M$ and in the second step to rebroadcast the time-reversed signal $\eta\left(x_M,-t\right)$ from this unique point in order to observe the solution $\eta\left(x,-t\right)$ in the whole medium. The time-reversal approach has been shown to work in the linear regime for water waves \cite{Fink5}. However, whether time-reversal will generate breather solutions, which are fundamentally based on nonlinear interactions, under experimental conditions in a water wave tank is an open question, which is addressed in this study. 

In order to experimentally demonstrate the time-reversal refocusing of a considered doubly-localized rational NLS breather in water waves, the experiments is performed in an unidirectional flume, for two different doubly-localized breather solutions, following five steps. First, we start the experiment, by generating from an unique source (a single flap located at $x_S$) the temporal modulation of the breather at its maximal compression in water, i.e. $\eta\left(x=x_S,t\right)$, see Eq. (\ref{se}). As described by NLS theory, the wave field radiated by this source will propagate and decrease in amplitude during its propagation along the wave flume. A single wave gauge, placed at another specific position $x_M$ far from the wave maker collects this attenuated wave profile. In a third step, the latter signal is reversed in time, providing therefore new initial conditions to the wave generator and initiating the next stage, which consists in generating the time-reversed attenuated wave signal. If the time-reversal symmetry is valid, one should expect the refocusing and the perfect reconstruction of the initial maximal breather compression, measured at the unchanged wave gauge position $x_M$ as last step, taking into account the complex mixture between all sinusoidal frequencies and phase components during the complex modulation instability process. Next, we describe the experimental setup, the performed experiments as well as the obtained results. 

\begin{figure}[h]
\centering
\includegraphics[width=\columnwidth]{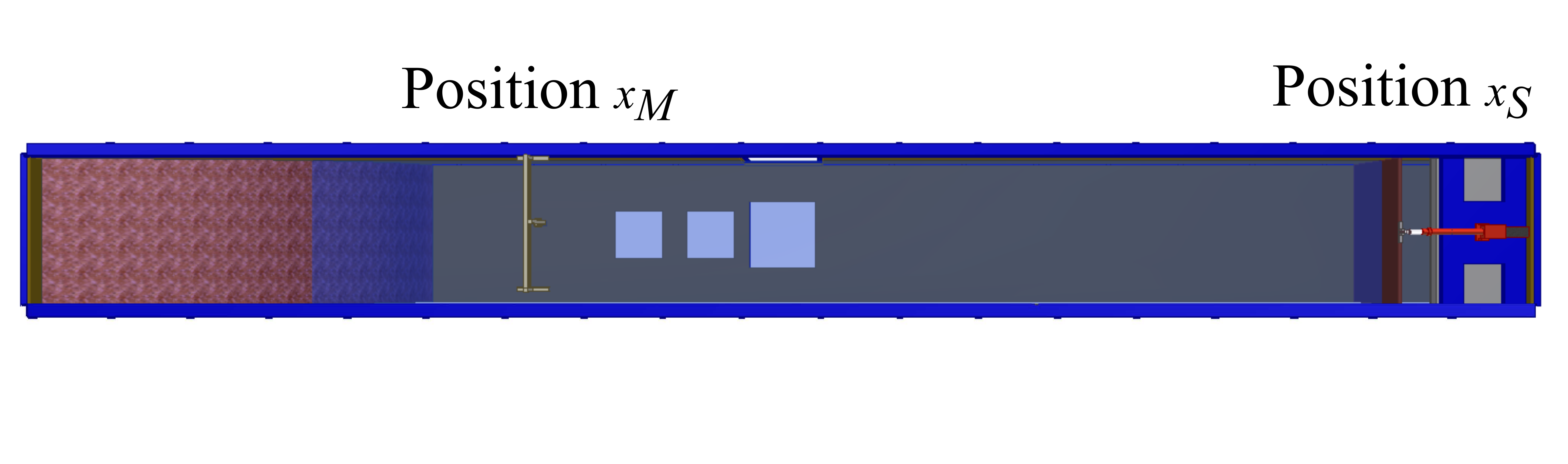}
\caption{Schematic upper view illustration
 of the wave
 flume. The single flap, driven by a hydraulic cylinder, is installed at the right end of the wave flume at "Position $x_S$". The wave gauge is placed 9 m from the flap, which location is labeled by "Position $x_M$". The gauge is far away from the absorbing beach, displayed at the left end of the flume.} \label{fig2}
\end{figure}

Experiments have been conducted in an unidirectional 15 m long wave flume with a constant water depth of $h_0=1$ m. The surface gravity Stokes waves are generated by a single flap, installed at one end of the flume, labeled by "Position $x_S$", and is driven by a computer-controlled hydraulic cylinder. At the other end, an in the water submerged absorbing beach is installed in order to avoid wave reflections. In the whole set of the performed experiments, the capacitance wave gauge, which measures the surface elevation of the water with a sampling frequency of 500 Hz, is placed 9 m from the flap and is {\it fixed} at this position, denoted by "Position $x_M$". This position has been determined to satisfy a reasonable propagation distance in the wave flume, while the wave gauge is still separated by 3 m from the installed beach to prevent noticeable wave reflection effects. The whole experimental setup is illustrated in Fig. \ref{fig2}

It is important to mention that ideal experimental conditions should be provided to minimize the dissipation effects, which have a strong influence on the soliton propagation during its evolution in a water wave tank. The walls of the flume were therefore properly cleaned and the water was filtered accordingly before performing the experiments. 
														
First, we generate the wave profiles of the breathers, satisfying Eq. (\ref{nls}), i.e., in dimensional units, at their maximal amplitude amplification. Fig. 3 shows these initial wave profiles, which amplitudes are amplified by a factor of three for the Peregrine and of five for the Akhmediev-Peregrine solution, respectively. 

\begin{figure}[h]
\centering
\includegraphics[width=\columnwidth]{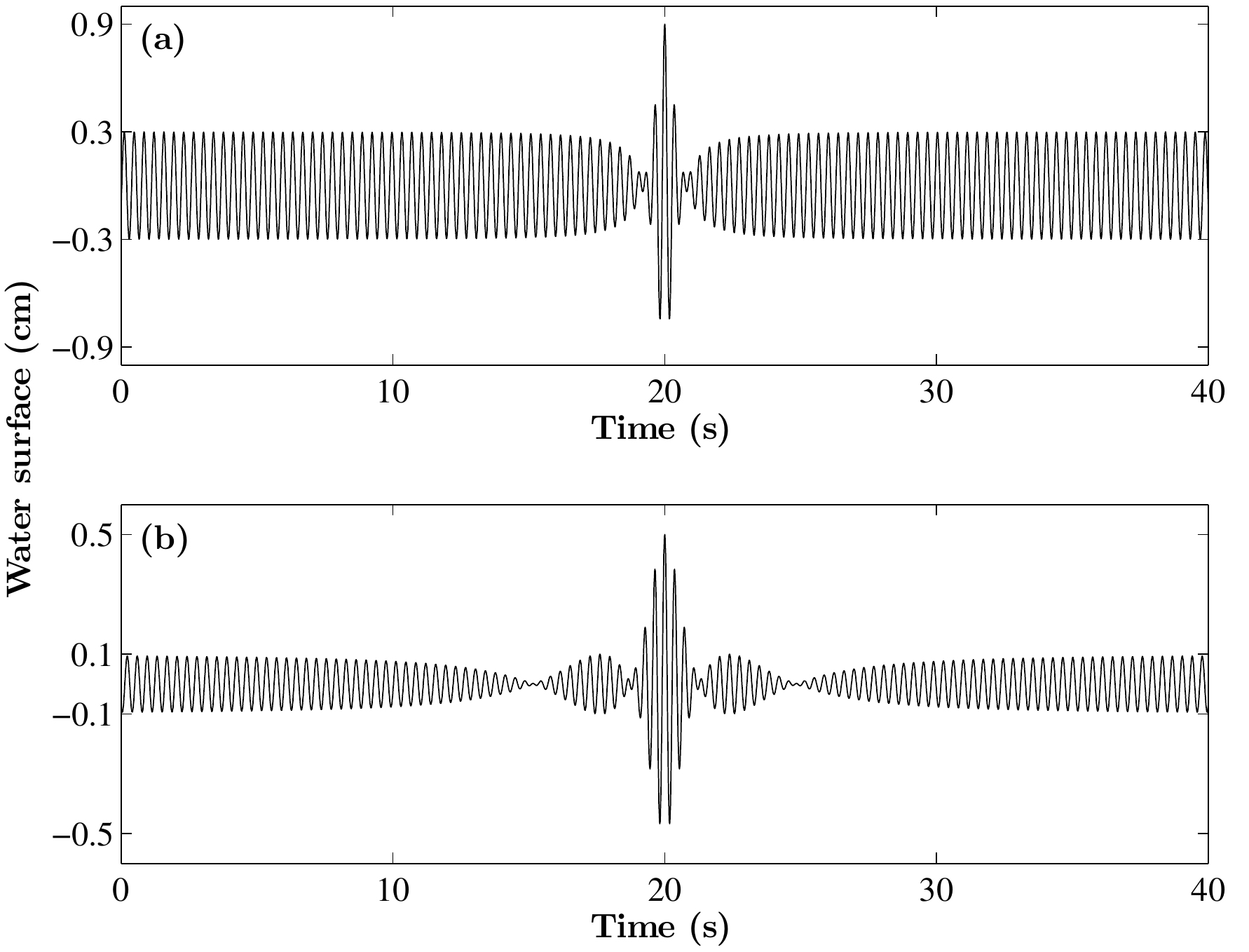}
\caption{(Color online) Initial conditions provided by theory Eq. (\ref{se}) at the position of maximal amplification, i.e., at $X=0$, applied to the wave paddle at Position $x_S$: (a) the Peregrine water wave profile for the carrier parameters: $a_0=0.3$ cm and $\varepsilon_0=0.09$, (b) the Akhmediev-Peregrine water wave profile for the carrier parameters: $a_0=0.1$ cm and $\varepsilon_0=0.03$.}\label{fig3}
\end{figure}

\begin{figure}[h]
\centering
\includegraphics[width=\columnwidth]{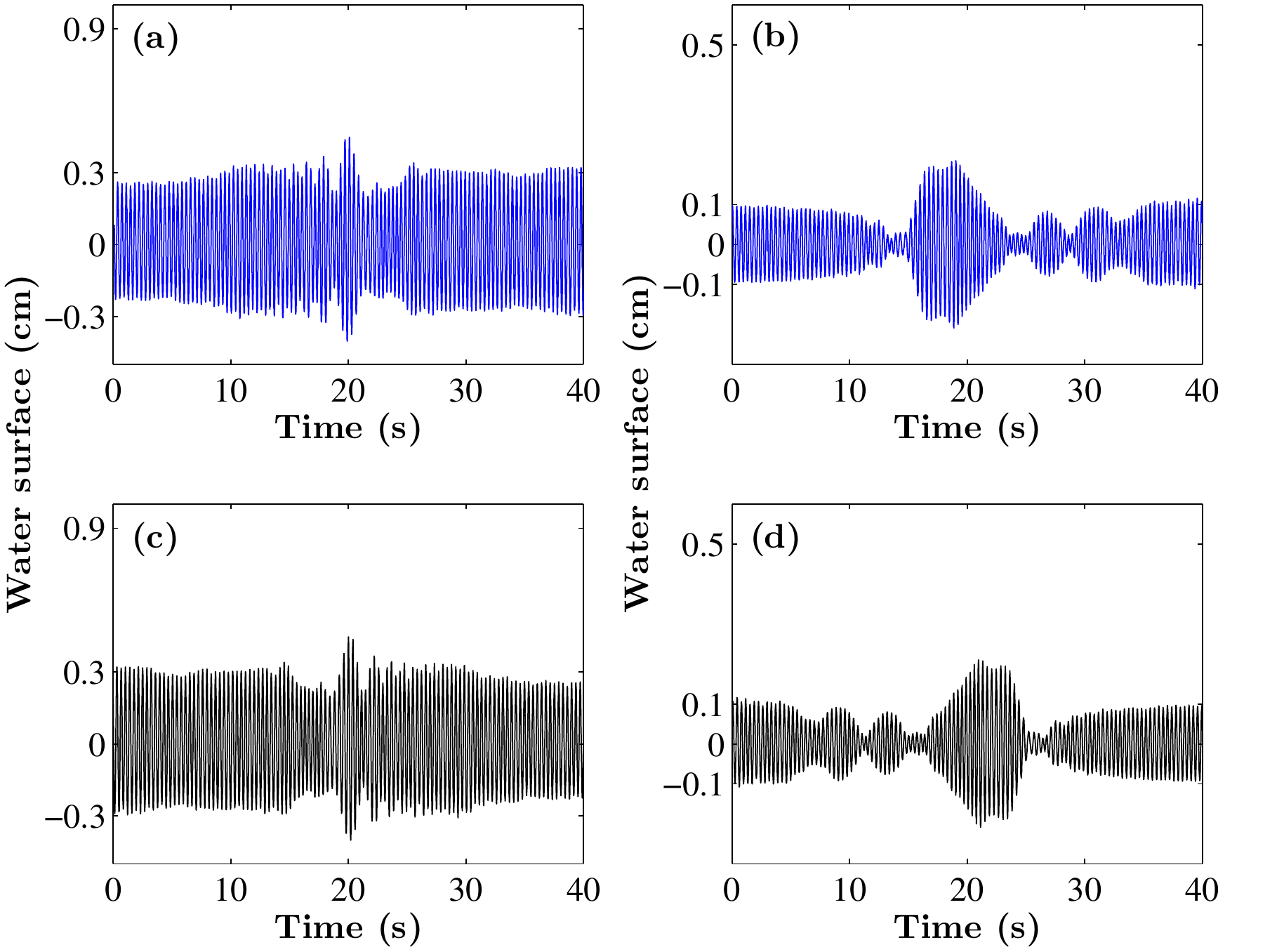}
\caption{(Color online) (a) Surface water wave profile of the in amplitude decreased Peregrine breather, measured 9 m from the flap at Position $x_M$. (b) Surface water wave profile of the in amplitude decreased Akhmediev-Peregrine breather, measured 9 m from the flap at Position $x_M$. (c) Time-reversed signal of the time-series shown in (a) providing new initial conditions to the flap and reemitted at Position $x_S$. (d) Time-reversed signal of the time-series shown in (b) providing new initial conditions to the flap and reemitted at Position $x_S$.}\label{fig4}
\end{figure}

At that initial stage of maximal breather compression, the focused high-amplitude waves are obviously strongly nonlinear. It is crucial for the experiments to avoid initial wave breaking in the wave flume. Consequently, the carrier wave parameters, determined by the amplitude $a_0$ and the steepness $\varepsilon_0:=a_0k_0$, have to be chosen accordingly and are set to be $a_0=0.3$ cm and $\varepsilon_0=0.09$ for the Peregrine as well as $a_0=0.1$ cm and $\varepsilon_0=0.03$ for the Akhmediev-Peregrine solution. The chosen steepness values are far from the experimentally determined wave breaking thresholds \cite{ChabchoubPRE,Babanin}. Knowing the amplitude $a_0$ and the steepness $\varepsilon_0$ of the background, the wave number is trivially $k_0=\frac{\displaystyle\varepsilon_0}{\displaystyle a_0}$. The wave frequency can then be determined from the linear dispersion relation of waves in deep-water $\omega_0=\sqrt{gk_0}$, where $g=9.81\operatorname{m}\cdot\operatorname{s}^{-2}$ denotes the gravitational acceleration. It has been previously shown that within the range of chosen amplitudes, the response function of the flap can be assumed to be linear, while surface wave profiles are generated with high accuracy \cite{ChabchoubPRE}. Fig. \ref{fig3} shows the initial conditions applied to the flap at the Position $x_S$. 
As next step, after generating the Peregrine and the Akhmediev-Peregrine breather at their maximal wave amplitude of 0.9 cm and 0.5 cm, respectively, we collect the wave profiles after having being declined in amplitude, as predicted by theory, with the wave gauge at the Position $x_M$, i.e., 9 m from the flap Position $x_S$. The corresponding data are shown in Fig. \ref{fig4} (a) and (b). 

\begin{figure}[h]
\centering
\includegraphics[width=\columnwidth]{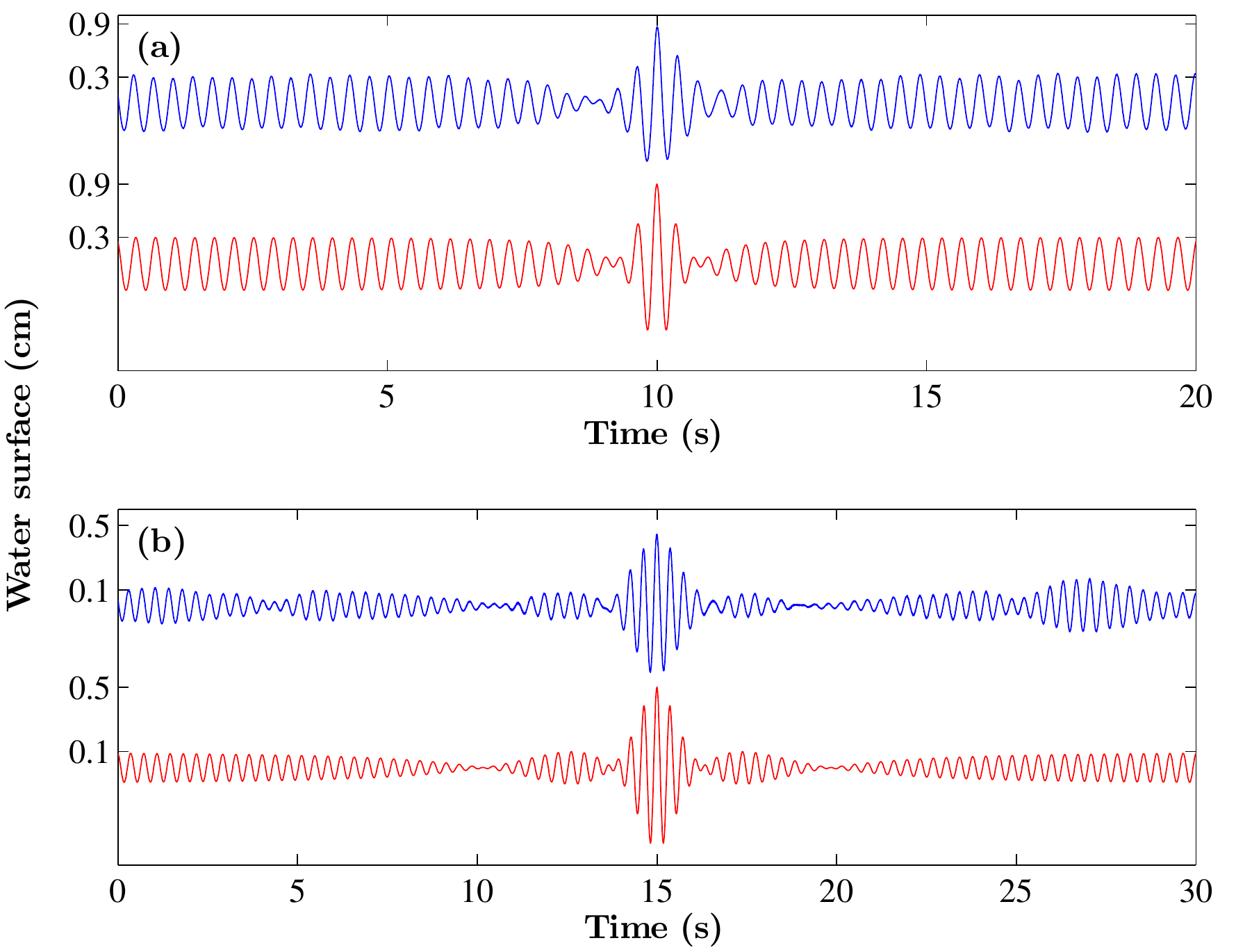}
\caption{(Color online) (a) Comparison of the Peregrine surface profile measured 9 m from the paddle at Position $x_M$, starting its evolution from time-reversed initial conditions (blue upper line) with the expected theoretical NLS wave profile at the same position (red bottom line). (b) Comparison of the Akhmediev-Peregrine surface profile measured 9 m from the paddle at Position $x_M$, starting its evolution from time-reversed initial conditions (blue upper line) with the expected theoretical NLS wave profile at the same position (red bottom line).}\label{fig5}
\end{figure}

The third step consists in reversing these recorded breather signals in time. The latter time-reversed signals, illustrated in Fig. \ref{fig4} (c) and (d), provide now new initial conditions for the wave maker in order to initiate the fourth stage of the experiment. If the breather dynamics is time-reversal invariant, it is expected to observe the refocusing of the waves after reemitting the latter small modulated time-reversed signals to the single flap. Note that due to the spatial reciprocity of the NLS equation we can reemit the time-reversed signal from the point $x_S$ and observe the refocusing at point $x_M$, instead of rebroadcasting the reversed wave field from $x_M$ and expecting refocusing at the position $x_S$. At the fifth and last step of the experiments, we measure the surface elevations related to the time-reversed initial conditions, again 9 m from the flap at the Position $x_M$. The corresponding measurements are illustrated in Fig. \ref{fig5}.

Clearly, Fig. \ref{fig5} shows the accurate refocusing and reconstruction of the breather surface elevation of the corresponding NLS solution, already presented in Fig. \ref{fig3}. The results are in a very good agreement with the theoretical predictions, expected at this position within the framework of NLS hydrodynamics: maximal amplitude surface water waves is of 0.9 cm for the Peregrine breather and of 0.5 cm for the Akhmediev-Peregrine breather, which correspond to values of amplification of amplitude, related to the corresponding NLS solution at the maximal stage of breather compression, as generated in the first step of the experiment.  Generally, comparisons of surface wave profiles with theory should be conducted by taking into account the influence of bound waves \cite{ChabchoubPRX,ChabchoubPRE}. However, due to the small steepness values, chosen for the analysis, the contribution of the higher Stokes harmonics is small and the main wave dynamics can be described by Eq. (\ref{se}). The time scale is adapted to the chosen steepness values. Since the investigated doubly-localized solutions describe asymptotically the case of infinite wave modulation, the non-modulated regular wave field of almost constant amplitude should be mapped within the time scale. These observations prove the time-reversal invariance of the NLS and especially, of strongly nonlinear water waves. In addition, the results endorse the accuracy of the NLS, describing the complex evolution dynamics of non-breaking hydrodynamic rogue waves. Some discrepancies between theory and experiment can be also noted with respect to the shape of the wave profiles, as can be seen in Fig. 5. The latter are due to higher-order nonlinearities (Stokes effects) and to higher-order dispersion, not taken into account in the NLS approach, as well as to occurring experimental imperfections, including dissipation and wave reflection, naturally existing while conducting the experiments. Nevertheless, the experiments are a clear confirmation of the possibility to reconstruct strongly localized, thus, strongly nonlinear waves through time-reversal. This confirms the fact that this technique can be used to construct new time-reversal invariant localized solutions of nonlinear evolution equations, which considerably amplify the amplitude of a wave field, therefore, also in the case of strong nonlinearity. 

To summarize, we have shown that doubly-localized NLS breather solutions can be experimentally reconstructed using the time-reversal technique. This confirms the time-reversal invariance of the NLS and underlines again its accuracy to describe the propagation dynamics of water waves. The results suggest that strongly nonlinear localized waves may be experimentally reconstructed in other nonlinear dispersive media governed by the NLS, such as optics, Bose-Einstein condensates and plasma, using this method. Potential applications of time-reversal would range from remote sensing to rogue wave time-series analysis in order to reconstruct the dynamics of rogue waves, already formed in the oceans or in fibers. It is obvious that this method may be also applied to predict and better understand the sudden occurrence of extreme wave events, still considered to be mysterious. Numerical simulations, based on a higher-order nonlinear approach, such as the Dysthe-type model \cite{Trulsen}, or based on fully nonlinear simulations \cite{Slunyaev}, may characterize the applicability limitations and restrictions to the time-reversal principle in the evolution of NLS-type rogue waves as well as of strongly nonlinear waves in complex media.

A.C. acknowledges partial support from the Australian Research Council (Discovery Projects DP1093349 and DP1093517). This work was supported by LABEX WIFI (Laboratory of Excellence ANR-10-LABX-24) within the French Program “Investments for the Future” under reference ANR-10-IDEX-0001-02 PSL$^*$.

\end{document}